\documentclass{aa}
\usepackage{xcolor}
\usepackage{graphicx}
\usepackage[fleqn]{amsmath}	
\usepackage{amssymb}	
\usepackage{mathtools}
\usepackage{txfonts}
\usepackage{hyperref}
\hypersetup{
  colorlinks=true,   
  citecolor=blue,    
  urlcolor=blue,     
  linkcolor=blue,     
}

\newcommand{\lo}{L_{\rm UV}}
\newcommand{\lx}{L_{\rm X}}
\newcommand{\fo}{F_{\rm UV}}
\newcommand{\fx}{F_{\rm X}}
\newcommand{\om}{\Omega_{\rm M}}
\newcommand{\ol}{\Omega_\Lambda}
\newcommand{\lcdm}{$\Lambda$CDM}

\newcommand{\gammax}{\Gamma_{\rm X}}

\begin{document} 

   \title{The X-ray-to-UV relation does not evolve in homogeneous quasar samples}
   
   \author{
    G. Risaliti\inst{1,2}\thanks{\email{guido.risaliti@unifi.it}}, E. Lusso\inst{1,2}, E. Nardini\inst{2}, C. Niccolai\inst{3,4}, M. Ralowski\inst{1,5,6}, A. Sacchi\inst{7},
A. Shenstlova\inst{1,8}, M. Signorini\inst{2,9}, B. Trefoloni\inst{2,10}  }

\institute{
$^{1}$ Dipartimento di Fisica e Astronomia, Universit\`{a} di Firenze, via G. Sansone 1, 50019 Sesto Fiorentino, Firenze, Italy\\
$^{2}$ INAF-Osservatorio Astrofisico di Arcetri, Largo E. Fermi 5, 50125, Firenze, Italy\\
$^{3}$ Scuola Superiore Meridionale, Via Mezzocannone 4, I-80138 Napoli, Italy\\
$^{4}$ Istituto Nazionale di Fisica Nucleare, sez. di Napoli, Via Cinthia 9, I-80126 Napoli, Italy\\
$^{5}$ Astronomical Observatory of the Jagiellonian University, Faculty of Physics, Astronomy and Applied Computer Science, ul. Orla 171, 30-244 Cracow, Poland\\
$^{6}$ Jagiellonian University, Doctoral School of Exact and Natural Sciences, Astronomy, Cracow, Poland\\
$^{7}$ INAF–Istituto di Astrofisica Spaziale e Fisica Cosmica di Milano, via A. Corti 12, I-20133 Milano, Italy\\
$^{8}$ Instituto de Astrofísica, Facultad de Física, Pontificia Universidad Católica de Chile, Casilla 306, Santiago 22, Chile\\
$^{9}$ European Space Agency (ESA), European Space Research and Technology Centre (ESTEC), Keplerlaan 1, 2201 AZ Noordwijk, The Netherlands\\
$^{10}$ Scuola Normale Superiore, Piazza dei Cavalieri 7, I-56126 Pisa, Italy
}
   \date{\today}

 
  \abstract{
  We present a new, highly homogeneous quasar sample with X-ray and UV observations optimized to reliably estimate distances via the non-linear X-ray-to-UV relation. Cross-matching the Sloan Digital Sky Survey DR16 quasar catalog with the XMM-{\it Newton} serendipitous catalogue (4XMM--DR14), we employ strict selection criteria to build a robust sample: (1) UV and (2) X-ray colour constraints to avoid, respectively, extinction and absorption; (3) removal of broad absorption line and radio-bright quasars; (4) exclusion of sources at $z$\,$<$\,0.7 to prevent galactic UV contamination; and (5) rejection of sources with shallow X-ray observations. The latter step, closely related to the Eddington bias, is critical because SDSS data are generally deeper than X-ray data for typical quasar spectral energy distributions: ignoring such a discrepancy introduces a spurious redshift dependence in the X-ray-to-UV relation parameters. Our final sample contains $\sim$2,000 quasars at $z$\,=\,0.7--5. We demonstrate that the X-ray-to-UV relation is constant across this redshift range, with a mean slope of 0.58\,$\pm$\,0.01 and a dispersion of 0.15 dex. 
  Our findings confirm the intrinsic stability of this relation over cosmic time, emphasizing that both homogeneity and robust Eddington bias corrections are vital for flux-limited samples. 
  In fact, the impact of the preferential detection of X-ray brighter-than-average sources near the effective sensitivity limits significantly grows with redshift. Any resulting evolutionary trend in the X-ray-to-UV relation, especially in the form a slope flattening, is therefore either a largely spurious effect, or the result of mixing populations of quasars with intrinsically different spectral properties. 
  }

\titlerunning{A new quasar sample for cosmology}
\authorrunning{G. Risaliti et al.}
   \maketitle
%

\section{Introduction}
Accreting super massive black holes (SMBHs) in the centre of galaxies (i.e., active galactic nuclei, AGN hereafter, of which quasars represent the brightest members) show characteristic observational features over a wide range of wavelengths. The spectral energy distribution (SED) of unobscured AGN present a rising emission towards the UV, which forms the so-called big blue bump, and a softening (flattening) at shorter wavelengths \citep[see, e.g.,][and references therein]{saccheo2023}, which originates in an optically thick disc surrounding the SMBH \citep{1973A&A....24..337S,1973blho.conf..343N}. At X-ray energies ($>$\,0.2--0.5 keV), the main continuum component is well described by a power law with a photon index $\gammax$\,$\simeq$\,2 up to energies of a few hundreds of keV, and the X-ray photons are produced by inverse-Compton scattering of disc UV photons by a hot plasma of relativistic electrons (the so-called X-ray corona; \citealt{HM91}). 

A key observational result on the connection between disc and coronal emission is provided by the non-linear relation between the monochromatic UV luminosity at 2500 \AA\ ($\lo$) and the one in the X-rays at 2 keV ($\lx$; \citealt{avnitananbaum79,zamorani81,avnitananbaum86}).
This relation, parametrized as $\log\lx$\,=\,$ \gamma\log\lo$\,+\,$\beta$, has been observed in both optically and X-ray selected AGN samples, with a slope $\gamma$\,$\simeq$\,0.6 \citep[e.g.][]{vignali03,strateva05,steffen06,just07,green09,lusso2010,young2010,2012A&A...539A..48M,2012MNRAS.422.3268J}, implying that a tenfold increase in UV luminosity yields only a fourfold increase in X-ray luminosity. Thus, optically bright AGN emit relatively less X-rays than optically faint AGN.
The typical scatter observed in this relation for the overall AGN population is on the order of 0.4 dex. Yet, when the AGN sample is selected to be homogeneous, adopting strict selection criteria such as, for instance, UV and X-ray colour constraints to avoid extinction and absorption by dust and gas, and the removal of broad absorption line (BAL) and radio-bright quasars, the observed dispersion reduces to 0.24 dex \citep[][LR16 hereafter]{lr16}, or even lower when variability is also taken into account \citep[i.e., $\simeq$\,0.1 dex;][]{signorini2024}. The tightness of the $\lx$--$\lo$ relation, when homogenous samples of blue bright AGN are selected, is observed over a wide redshift range, from local objects up to $z$\,$\simeq$\,6 (see also \citealt{vito2019}), suggesting that the physical mechanism driving the UV and X-ray emission is universal across different cosmic epochs. 
Therefore, the $\lx$--$\lo$ relation has emerged as a powerful tool, for both advancing our physical understanding of quasars \citep[see, e.g.,][and references therein]{kang2025,kammoun2025} and their applicability to cosmology \citep[e.g.,][]{rl19}.

A key point is whether the $\lx$--$\lo$ relation evolves with redshift.
The vast majority of previous works in the literature found no observational evidence in favour of a redshift evolution \citep[e.g.,][]{just07,steffen06,lusso2010,lr16,timlin2021}. Yet, a few recent studies report a trend with redshift of the $\lx$--$\lo$ correlation parameters \citep[see][and references therein]{rankine2024,chira2026}, which would have significant consequences not only on accretion physics but also on the cosmological application of the relation.

Regarding the latter point, in \citet{lusso2025} we recently reviewed the main steps necessary to obtain reliable AGN distance estimates and effectively use them to construct a high-redshift Hubble diagram. 
Briefly, the key requirements to build a homogeneous AGN sample at both optical/UV and X-ray energies are the following: firstly, we must obtain reliable flux measurements, both in the X-ray and UV bands. In particular, dust-reddened and gas-obscured objects should be neglected to accurately estimate the intrinsic continuum emission. 
In principle, a detailed correction of reddening/absorption effects would require a complete spectral analysis of both the (rest-frame) UV and X-ray spectra of each source. In practice, however, our strategy aims at finding the optimal trade-off between the precision of the flux measurements and the size of the sample.
Secondly, the sample selection must take into account possible biases due to the flux limits of the parent UV and X-ray catalogues. This is a particularly critical point, considering that such biases are expected to be redshift-dependent, due to the fact that the fraction of objects close or below the flux limit increases with redshift (we refer to this effect as to the `Eddington bias'). Finally, it must be verified that the selected sample is characterized by an X-ray-to-UV relation that does not evolve with redshift (in flux--flux space). With this respect, 
we demonstrated that both the slope $\gamma$ and the intercept $\beta$ of the $\lx$--$\lo$ relation do not evolve with redshift, provided that a proper selection scheme is enforced (e.g., \citealt{lusso2020,lusso2025}). This is supported by direct observational evidence showing a constant $\gamma$ for the $\fx$--$\fo$ relation within narrow redshift bins up to $z$\,$\simeq$\,3.5, and indirectly for $\beta$ through a comparison of the AGN and Type Ia supernovae Hubble diagrams in their common redshift range (i.e., up to redshift $z$\,$\simeq$\,1.5).
The fact that there is no direct way to observationally prove that the intercept of the relation does not evolve with redshift is an intrinsic limitation of the cosmological application: the redshift independence must be assumed based on physical arguments and/or on the agreement of the distance estimates with other probes. 

Our primary objective is to confirm that the X-ray-to-UV relation remains tight and with a constant slope across redshift. To achieve this, we present a new, highly homogeneous, optically selected AGN sample with X-ray and UV observations optimized for reliable distance measurements. Using this sample, we show that previously reported evolutionary trends or slope flattening are largely spurious, being entirely attributable to an incomplete correction for the Eddington bias, which implies that a growing fraction of X-ray detections at high redshift consists of sources preferentially caught in a high flux state, therefore introducing a bias.

Source luminosities are estimated by adopting a concordance flat $\Lambda$CDM cosmology with $H_0$\,=\,70 km s$^{-1}$ Mpc$^{-1}$, $\om$\,=\,0.3, and $\ol$\,=\,0.7.

\section{Sample Selection}
\label{Sample Selection}
The AGN sample is selected from the DR16 catalogue of the Sloan Digital Sky Survey (SDSS; \citealt{wu2022}), which contains 750,414 spectroscopically confirmed AGN in the redshift range 0.01\,$\leq$\,$z$\,$\leq$\,7.01. Following our approach to select homogeneous AGN properties at both UV and X-ray energies (see LR16; \citealt{lusso2020,lusso2025} for details), we removed from this catalogue all the sources flagged as BALs (where sources with BAL\_PROB\,=\,0 are considered non-BALs) and all AGN 
flagged as radio bright by \citet[][amounting to $\simeq$\,8\% of the main SDSS sample]{mingo2016}. This yields 623,810 SDSS quasars. We further excluded 889 quasars classified as BALs by \citet{2009ApJ...692..758G} and 1408 AGN that have a radio flux entry in the FIRST catalogue (which contains 946,932 observations up 14 December 2017) within 5$^{\prime\prime}$ from the SDSS position. The final parent SDSS AGN sample is composed by 621,511 sources.\footnote{Two AGN have redshift lower than zero and have been thus excluded from the sample.}

This SDSS AGN sample is then cross-matched with the XMM-{\it Newton} source catalogue 4XMM--DR14 \citep{webb2020}. 
4XMM--DR14 is the fourth generation catalogue of serendipitous X-ray sources, which contains 1,035,832 detections (692,109 unique X-ray sources) made publicly available by 2023 December 31.\footnote{http://xmmssc.irap.omp.eu/Catalogue/4XMM-DR14/4XMM\_DR14.html} 
The net sky area covered (taking into account overlaps between observations) is $\sim$1383 deg$^2$, for a net exposure time of $\geq$\,1 ks.

To select reliable X-ray detections, we have applied the following quality cuts to the 4XMM--DR14 catalogue: SUM\_FLAG\,$<$\,3 (low level of spurious detections), OBS\_CLASS\,$\leq$\,3 (quality classification of the whole observation)\footnote{For more details the reader should refer to the 4XMM catalogue user guide.}, EP\_TIME\,$>$\,0 (EPIC exposure time available) and CONFUSED\,=\,false (indicating that a specific X-ray detection has a probability of being associated with two or more distinct, overlapping X-ray sources). These filters lead to 865,935 X-ray detections.
We have adopted a maximum separation of 3$^{\prime\prime}$ to provide optical classification and spectroscopic redshift for all the cross-matched objects. 
This yields 35,090 XMM-{\it Newton} observations: 21,655 unique sources (6,238 of which have $\geq$\,2 observations) covering the redshift range 0.053\,$<$\,$z$\,$<$\,7.007. 

For each XMM-{\it Newton} detection, we have computed the EPIC sensitivity ($5\sigma$ minimum detectable flux) at 2 keV following the same approach as in \citet[][see also LR16]{lusso2020}, which we then use to define the limiting flux employed for the Eddington bias correction. We describe this process in detail in Section~\ref{Depth of the X-ray observations}. To select only unobscured XMM-{\it Newton} sources, we computed the X-ray photon index from the broadband X-ray flux photometry following a procedure similar to that described in \citet{lusso2020}, to which we made several improvements as described in Section~\ref{X-ray slope}. 
We have then applied the following filters to each X-ray detection: we selected (1) all detections with an X-ray spectral slope typical of unobscured AGN; and (2) all X-ray observations deep enough to ensure the removal of the Eddington bias towards X-ray brighter-than-average sources (see Section~\ref{Depth of the X-ray observations}). Finally, following the results presented by LR16 (see their Section~4) and \citet{signorini2024}, we averaged all X-ray observations for sources with multiple detections that meet the selection cuts above, thus reducing the effect of X-ray variability on the dispersion of the $\lx$--$\lo$ relation. 

A further filter applied to the sample concerns the selection of gas/dust-unobscured AGN at optical/UV wavelengths, as severe line-of-sight attenuation requires a highly uncertain correction of the observed fluxes.
To do so, we built the broadband SED from the near-IR to the UV for all the AGN in the sample, and we selected all sources with optical/UV colours similar to those of a stacked spectrum of blue, unobscured quasars \citep[][see Section~\ref{Color selection}]{saccheo2023}. The rest-frame 2500-\AA\ continuum luminosities, $\lo$, are computed for each AGN from their photometric broadband SED following the same approach as described in Section 3 of \citet[][]{lusso2020}.

The core physics driving the non-linear $\lx$--$\lo$ relation relies on a `standard' radiative coupling between the accretion disc and the X-ray corona. The inclusion of sources where additional physical mechanisms are at play, such as relativistic jets in radio-loud quasars that bear a substantial non-thermal X-ray contribution, or powerful winds as in BAL quasars whose launch drains a sizeable fraction of the accretion power, significantly alter the observed fluxes. Failing to isolate a highly homogeneous AGN population introduces systematic shifts that can mimic (or obscure) a true redshift evolution. Therefore, our selection steps are designed to identify a sample where the observed emission at both UV and X-ray energies is nuclear (accretion)-dominated.

It is worth stressing that the selection criteria we employed are entirely decoupled from cosmological distances. To ensure our analysis remains strictly independent of any assumed cosmological model, we evaluated the impact of each filtering step on the $\lx$--$\lo$ relation within narrow redshift intervals. To fulfill the cosmology-independence requirement, the redshift bins must be sufficiently narrow that the relative difference in distance among sources within the each interval is negligible compared to the intrinsic dispersion of the $\lx$--$\lo$ relation. We have extensively discussed this strategy in previous works \citep[see, e.g.,][]{lusso2025}, demonstrating through simulations that any interval with $\Delta\log(z)$\,$<$\,0.1 is small enough to minimize any cosmology-dependent effects. In this study, we adopt an even more conservative approach by utilizing much tighter intervals with $\Delta\log(z)$\,=\,0.02, ensuring that cosmological effects are completely negligible. For presentation purposes, the results of the linear fits from these fine bins are subsequently averaged into larger redshift intervals to ease visualization.


We finally note that we could have built larger and more redshift-extended quasar samples by combining several different selections. These include, for example, SDSS--{\it Chandra} cross-matches, pointed X-ray observations at very high ($z$\,$>$\,4) and very low redshifts ($z$\,$<$\,0.5), and observations from specific surveys such as COSMOS \citep[e.g.,][]{bisogni2021}. Such an approach is the most powerful for building a Hubble diagram able to best constrain different cosmological models, and will be followed in a forthcoming paper (Lusso et al., in preparation). Here, however, we focus on a smaller but fully homogeneous sample, with the goal of keeping the error on the distances as small as possible to rigorously check for the absence of any redshift-dependent bias. 
Below, we describe in more detail the main steps adopted for the selection of this clean, homogeneous AGN sample.

\subsection{Optical/UV colour selection}
\label{Color selection}
\begin{figure}
\centering
\includegraphics[width=\linewidth,clip]{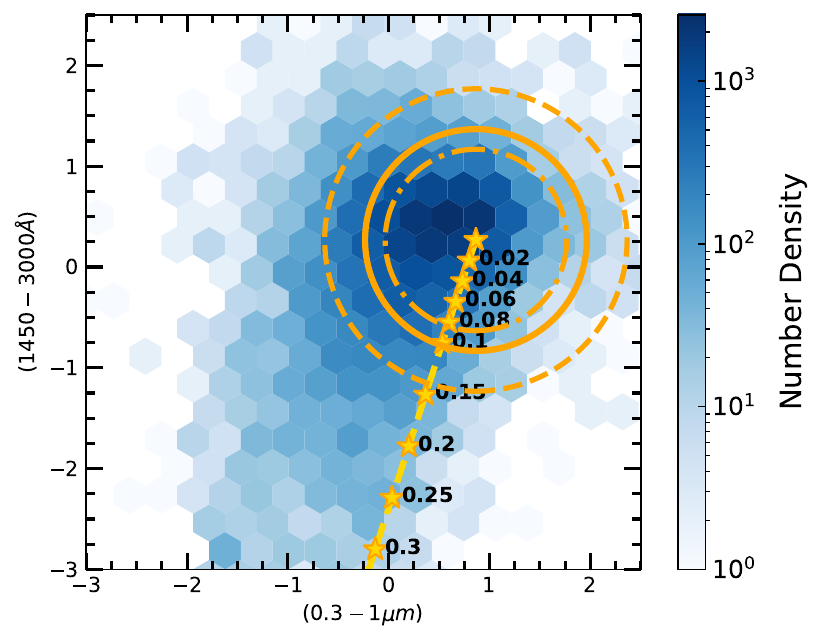}
\caption{Colour distribution for the SDSS--4XMM AGN sample. The plotted values represent the slopes of a power law in the $\log(\nu)$--$\log(\nu L_\nu)$ plane in the 0.3--1~$\mu$m and 1450--3000~\AA\ intervals, respectively. The solid line represents the value for the radius $R$\,=\,1.1, corresponding to a colour excess of $E(B-V)$\,$\sim$\,0.1. The dot-dashed and dashed lines represent radial values of 0.9 and 1.5, respectively. We selected all the AGN inside the circle defined by the solid line (see Section~\ref{Color selection} for details). The coloured bar represents the number density of AGN in each point. The stars represent the slope values of the quasar SED by \citet{saccheo2023} with increasing dust reddening (following the extinction law of \citealt{prevot84}), with $E(B-V)$ in the range 0--0.3. }
\label{fig:colourslopes}
\end{figure}

The colour selection at optical/UV wavelengths aims at avoiding both dust-reddened and host-galaxy contaminated objects, whose observed optical/UV continuum would require a model-dependent correction. Even if this correction did not introduce any systematic bias, it could substantially increase the scatter of the relation. We follow a procedure already tested and described in previous papers \citep[see, e.g., LR16;][for details]{rl19,lusso2020}, based on two optical colours (0.145--0.3~$\mu$m and 0.3--1.0~$\mu$m) derived from the rest-frame broadband SED, from optical to near-infrared wavelenghts. The resulting colours are illustrated in Figure~\ref{fig:colourslopes}. We select sources within a circle centred on the point corresponding to the colours of the average quasar SED published by \citet[][]{saccheo2023}, with a radius $R$\,=\,1.1, roughly equivalent to a colour excess $E(B-V)$\,$\simeq$\,0.12. In order to investigate the implications of this choice, we repeated the selection for different radii, as shown in Figure~\ref{fig:colourslopes} with dashed lines. Even if no significant differences in the overall $\lx$--$\lo$ relation were observed up to radii as large as $R$\,$\sim$\,1.5), we conservatively avoided radii $R$\,>\,1.1, for which the growing risk of including sources with underestimated UV fluxes outweighs the small improvement in sample size. We also checked that no change in the relation parameters is measured when choosing a smaller radius, such as $R$\,=\,0.9. In general, the weak dependence of the final outcome on the value of $R$ suggests that SDSS quasars are characterized by rest-frame blue colours by selection, and that the colour spread in Figure~\ref{fig:colourslopes} is possibly due to intrinsic SED differences rather than to dust reddening. We found a change in the relation parameters only if the few objects at $R$\,$>$\,1.5 are considered, likely because these sources are either genuinely reddened (towards the bottom left of the two-colour plane) or intrinsically peculiar. 

\begin{figure}
\centering
\includegraphics[width=1.05\linewidth,clip]{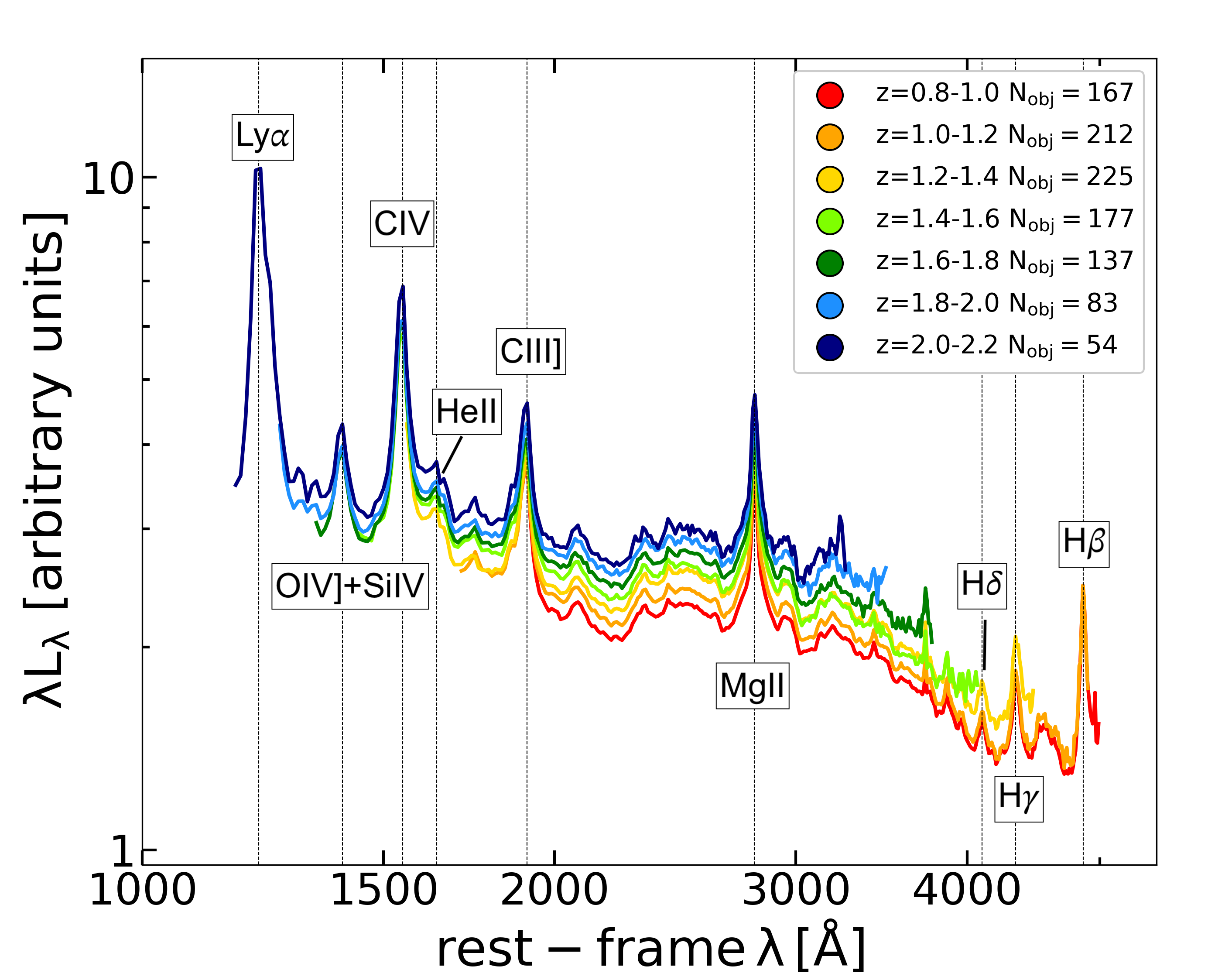}
\caption{Average SDSS spectra of our quasar sample in redshift bins, showing the absence of evolution of the optical/UV SED. Spectra are slightly shifted along the vertical axis for visualisation purposes.}
\label{fig:UVspectra}
\end{figure}

This colour selection is also aimed at selecting sources with minimal host-galaxy contamination. Yet, as pointed out in \citet{lusso2020}, some residual contamination from the host galaxy emission in the optical can still be present, especially for the low-redshift ($z$\,$\lesssim$\,0.7) and low-luminosity ($\lesssim$\,10$^{45}$ erg s$^{-1}$) AGN, for which the contrast between the nuclear and the host-galaxy emission is limited. Therefore, we conservately excluded from the sample all AGN at redshift $z$\,$\leq$\,0.7 \citep[see also][]{lusso2025}.

Our conservative selection criteria remove reddened and host-galaxy contaminated galaxies, significantly improving the homogeneity of the optical/UV spectral properties (more specifically, the continuum emission) in the AGN sample. 
To further test this, we randomly chose a sub-sample of about 1,000 AGN after the colour selection, and we computed the average composite of the SDSS spectra in redshift intervals where we have enough statistics (i.e., $N$\,$>$\,50 AGN in the redshift range 0.8--2.2). The resulting AGN composites are presented in Figure~\ref{fig:UVspectra}, where the absence of an evolution of the spectral properties across redshift, especially the continuum, is conspicuous.

\subsection{X-ray slope and luminosity}
\label{X-ray slope}
To compute the X-ray photon index, $\gammax$, and the rest-frame 2-keV luminosity, $\lx$, we followed anapproach similar to that described in \citet[][see also \citealt{rl19}]{lusso2020}. Below, we briefly highlight the main steps and the improvements compared to our previous works. 
Specifically, we considered the tabulated 0.5--2 keV (soft, $F_{\rm S}$) and 2--4.5 keV (hard, $F_{\rm H}$) fluxes reported in the 4XMM--DR14 serendipitous source catalogue. Differently from our previous works, we decided to employ a narrower hard X-ray band, given the low fraction of robust detections in 4.5--12 keV band: only 48\% of the 35,090 XMM detections have a signal-to-noise ratio $S/N$\,$>$\,1 at 4.5--12 keV, compared to 85\% at 2--4.5 keV.

The band-integrated fluxes were blueshifted to the rest-frame by considering pivot energy values of 1 keV ($E_{\rm S}$) and 2.8 keV ($E_{\rm H}$), respectively, and  assuming the same photon index used to derive the fluxes in the 4XMM catalogue (i.e., $\gammax$\,=\,1.42). For the soft band, the monochromatic flux at $E_{\rm S}$ is then:
\begin{equation}
\label{soft}
F_E(E_S)=F_{\rm S}\frac{(2-\gammax) E_{\rm S}^{1-\gammax}}{(2\,\rm keV)^{2-\gammax}-(0.5\,\rm keV)^{2-\gammax}},
\end{equation}
in units of erg s$^{-1}$ cm$^{-2}$ keV$^{-1}$. An equivalent expression holds for the hard band, with the obvious modifications. Flux values are corrected for Galactic absorption. 
A photometric photon index was then estimated as the slope of the power law connecting the two (soft and hard) monochromatic fluxes at the rest-frame energies corresponding to the observed pivot points. 
The rest-frame, photometric 2-keV flux (and its uncertainty) was interpolated (or extrapolated) based on such a power law. 
We then required each observation to provide a detection with a significance level of at least 2$\sigma$ in both bands.

The rationale for further filtering based on the X-ray slope mirrors that used for the optical colour: we aim to exclude potentially gas-obscured sources from our sample. This procedure inevitably removes intrinsically X-ray flat sources as well, thereby enhancing the spectral homogeneity of our sample also in the X-rays (cf. Figure~\ref{fig:UVspectra}). 
To determine the optimal trade-off between sample size and purity regarding X-ray obscuration, we tested several cuts on $\gammax$.
In this analysis, as with all subsequent ones, we test the impact of a specific filter while keeping all the other selection criteria active. For instance, here we examine the dependence on the X-ray slope for a sample that has already been filtered in terms of optical/UV colours and observation depth (see Section~\ref{Depth of the X-ray observations}). 
We apply a filter defined as $\gammax$\,--\,$\Delta(\gammax)$\,$>$\,$\Gamma_{\rm X,min}$, where $\Delta(\gammax)$ is the measured 1$\sigma$ error on the X-ray slope.
We analysed the effects of different thresholds, in steps of 0.05, starting from a value of $\Gamma_{\rm X,min}$\,=\,1 up to $\Gamma_{\rm X,min}$\,=\,2. 
We found that raising the threshold to a value of $\Gamma_{\rm X,min}$\,=\,1.8 yields a significant decrease in dispersion and a moderate, yet statistically significant, steepening of the $\lx$--$\lo$ relation. For higher values, we observe only a decrease in sample size without any further change in either the slope of the relation or its intrinsic dispersion.

\subsection{Depth of the X-ray observations}
\label{Depth of the X-ray observations}
The most critical step in our analysis addresses the Eddington bias inherent to flux-limited samples. This effect leads to the preferential detection of sources that are brighter-than-average in the X-rays when their expected flux is close to the observational detection limit. Because of the intrinsic dispersion in the $\lx$--$\lo$ relation, sources with `true' X-ray luminosities below the threshold can scatter above it due to statistical fluctuations or intrinsic variability. Crucially, since the high-redshift AGN population is observed almost entirely near the detection thresholds (unless a source benefits from a deep, pointed observation), this overestimation becomes progressively more severe with increasing redshift. If left uncorrected, this differential bias creates a spurious redshift dependence, artificially flattening the observed slope of the $\lx$--$\lo$ relation.


This bias is particularly detrimental to cosmological applications. As the fraction of sources observed near the flux limit naturally increases with redshift, an inadequate correction will systematically distort the $\lx$--$\lo$ relation, and consequently the derived distance estimates, in a redshift-dependent manner.
In all our works, including the present analysis, we have focussed exclusively on mitigating this bias within the X-ray observations. This is because our parent sample is drawn from the SDSS quasar catalogue, whose optical/UV data are, in almost all cases, significantly deeper than the accompanying serendipitous X-ray observations for a typical quasar.
Properly accounting for this effect is further complicated by the substantial intrinsic dispersion of the observed $\lx$--$\lo$ relation ($\delta$\,$\simeq$\,0.12--0.15 dex), which is primarily driven by intrinsic variability and orientation effects \citep{signorini2024}. 
To drastically reduce this selection bias, we must ensure that each X-ray observation is deep enough to detect a given source even if it undergoes a relatively large statistical fluctuation below its expected mean flux.
However, defining this threshold requires extreme care to avoid circularity with resepct to any cosmological inferences. 

The expected X-ray flux of an SDSS quasar can only be predicted via the $\lx$--$\lo$ \textit{luminosity} relation. Since the latter relation is inherently calibrated using luminosity distances, it depends directly on the  cosmological model used to convert redshifts into distances.
\label{Section: slope-redshift}
\begin{figure*}[!ht]
\centering
\includegraphics[width=0.7\linewidth,clip]{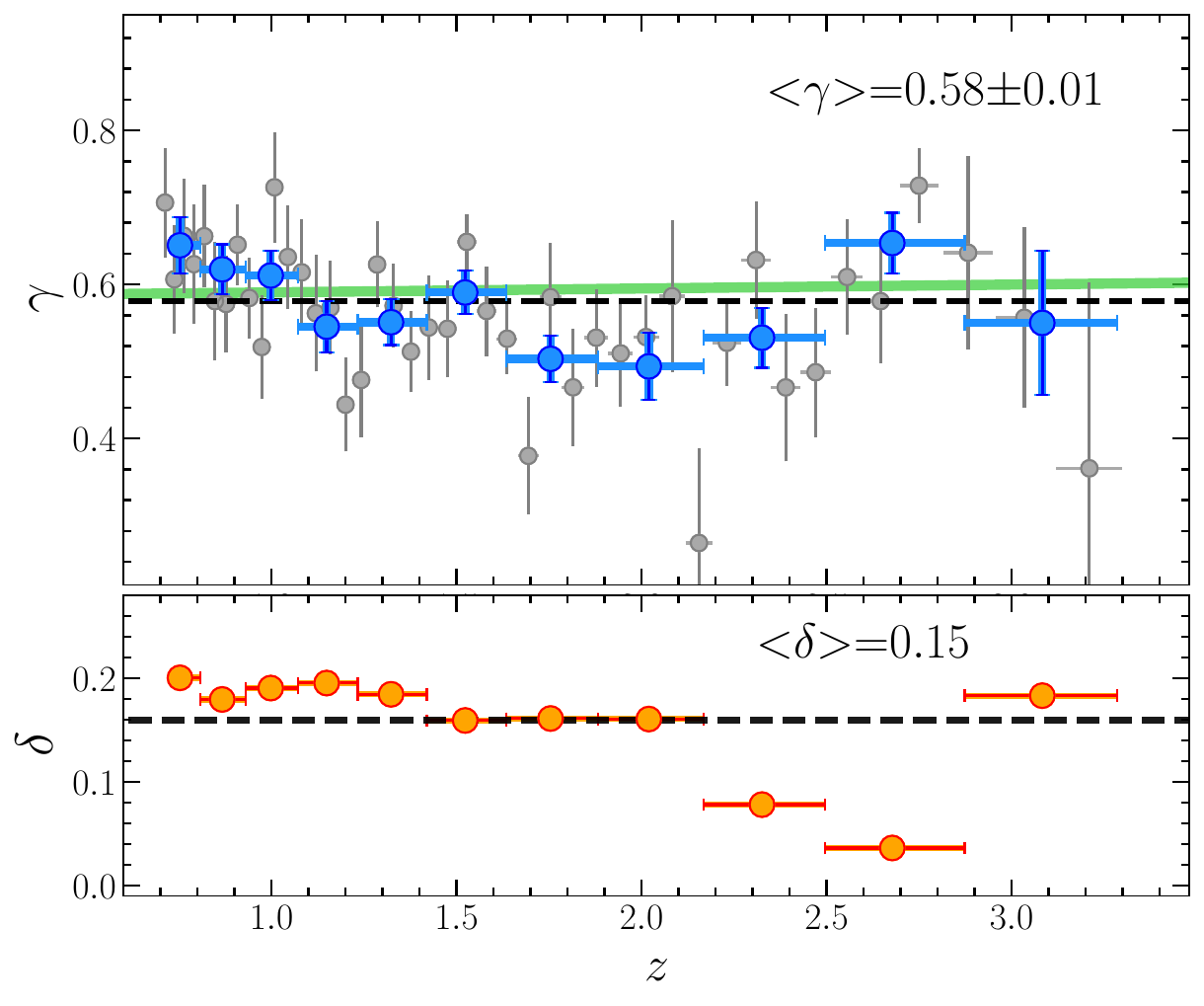}
\caption{{\it Top panel:} best-fit slope of the X-ray-to-UV relation in flux--flux space in narrow redshift bins ($\Delta(z)$\,=\,0.015, grey points). Blue points are obtained by merging  4 small intervals for ease of visualization. The dashed line marks the average $\langle\gamma\rangle$\,=\,0.58, while the solid line is the linear fit of the grey points, which demonstrates that the trend with redshift is remarkably flat and the intercept is statistically consistent with $\langle\gamma\rangle$ (see Section~\ref{Section: slope-redshift} for details). {\it Bottom panel:} intrinsic dispersion as a function of redshift. The dashed line marks the average value of $\langle\delta\rangle$\,$\simeq$\,0.15 dex.}
\label{fig:slopez}
\end{figure*}
To render any potential cosmological circularity negligible, we implemented the following multi-step procedure:
\begin{itemize}
    \item We estimated the expected X-ray flux based on the best-fit $\lx$--$\lo$ luminosity relation from \citet{lusso2020}. The slope is assumed to be $\gamma$\,=\,0.6, which implies an intercept $\beta$\,=\,8.2. Such a slope is fully consistent with the one derived in a cosmology-independent way from the average of $\gamma$ values determined from the $\fx$--$\fo$ relation in narrow redshift intervals, $\gamma$\,=\,0.586\,$\pm$\,0.061 \citep{lusso2020}). These baseline luminosities are derived assuming a standard, flat \lcdm\ cosmology with $\om$\,=\,0.3.
    \item We calculated luminosity distances ($D_L$) as a function of redshift across a broad range of alternative cosmological models. These include flat \lcdm\ configurations with varying matter densities (e.g., $\om$\,=\,0.1, 0.3, or 0.8) as well as $w$CDM dynamical dark-energy models (e.g., with $\om$\,=\,0.5, $w$\,=\,$-1.5$, or $\om$\,=\,0.2, $w$\,=\,$-0.5$). We then computed the maximum discrepancies among them. These variations naturally scale up with redshift, peaking at the highest redshift bounds of our sample ($z$\,=\,3--3.5). The maximum deviation between any given test cosmology and the concordance model is at most $\Delta\log(D_L) \simeq 0.15$ dex. 
    \item We defined a depth threshold for the X-ray observations ensuring a high source detection probability (at least $3\sigma$, see below) even under the combined conditions of an unfavorable test cosmology \textit{and} a severe negative statistical fluctuation of the X-ray flux. Given that the intrinsic dispersion of the $\lx$--$\lo$ relation can be as high as 0.15 dex, and that the contribution of cosmological model variations is also $\sim$\, 0.15 dex, 
    we conservatively adopt a value of 0.21 dex to represent the expected scatter of observed X-ray fluxes around the values predicted by the standard cosmology. 
    \item Based on the above considerations, we imposed a minimum required depth for each X-ray observation, defined as $\log F_{\rm X, min}$\,$<$\,$\log F_{\rm X, exp}$\,$-$\,$K$, where $K$ is a threshold parameter determined through empirical tuning. The flux limit $F_{\rm X, min}$ is evaluated as a function of off-axis angle and exposure time of each observation, following the methodology in \citet{lusso2020, lusso2025}. We tested a range of $K$ values, searching for the minimum threshold that yields a redshift-independent slope. Starting at $K$\,=\,0.45 (corresponding to a $\sim$\,2$\sigma$ downward fluctuation), we systematically increased the threshold in steps of $\Delta K$\,=\,0.05. As $K$ increases, the dispersion of the sample decreases and the average slope $\gamma$ increases until stabilizing at $K$\,$\sim$\,0.60--0.65. Higher values yield no further changes; we therefore adopt $K$\,=\,0.65.
    \item We performed detailed simulations to verify that this selection procedure successfully eliminates the Eddington bias. These simulations, presented in Appendix~A, validate the robustness of our approach, showing that the underlying cosmological model is recovered without bias for any $K$\,$>$\, 0.5. To maintain a highly conservative approach, we retain the stricter threshold of $K$\,=\,0.65 for the analysis of the real dataset.
\end{itemize}
The SDSS-4XMM AGN sample that meets all the selection criteria described above and corrected for the Eddington bias is composed by 1995 sources in the redshift range 0.7\,$<$\,$z$\,$<$\,5. 
The logarithmic UV luminosity ($\log \lo$) as a function of redshift for our final sample is shown in Figure~\ref{fig:loz}. Given the low number of sources available at high redshifts, our subsequent statistical analysis is restricted to the sub-sample at $z \le 3$, where the data density is sufficient to provide robust constraints.
\begin{figure}[h]
\centering
\includegraphics[width=\linewidth,clip]{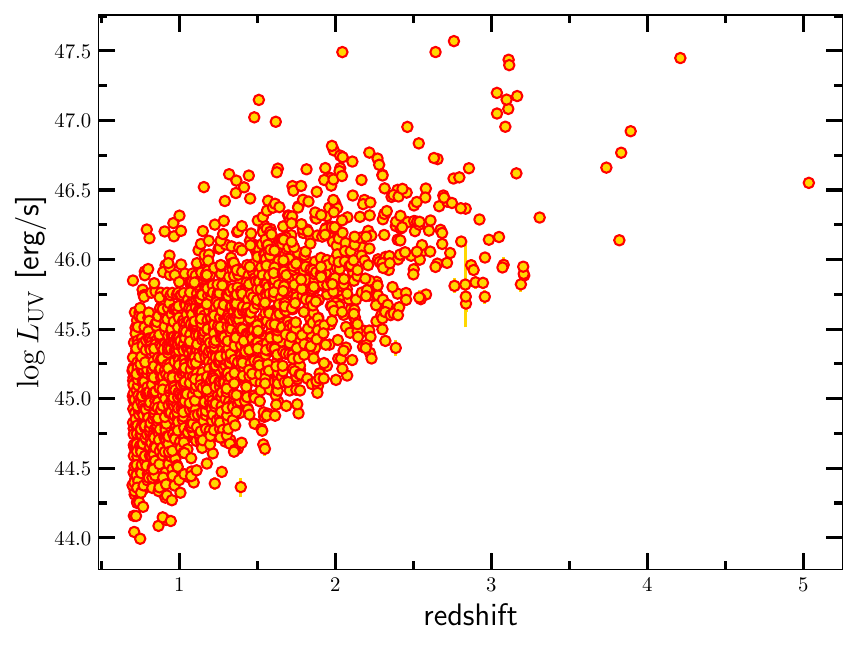}
\caption{Logarithmic UV luminosity ($\log \lo$, in units of erg~s$^{-1}$) as a function of redshift for the final clean SDSS-4XMM AGN sample.}
\label{fig:loz}
\end{figure}


A defining element of our strategy to construct an optimal sample for cosmological applications is that we do not prioritise maximising the total sample size; rather, we strictly isolate the population that best fulfills our rigorous physical requirements. This approach has two fundamental consequences. First, we deliberately prioritise individual distance precision and the mitigation of selection biases over complete sample statistics. Second, we do not require our sample to be representative of the broader, global quasar population. If our selection framework selectively favors a specific spectral subclass that exhibits enhanced reliability as a standardizable candle, we willingly restrict our analysis to that sub-population. 

Within a cosmological framework, the exclusion of optically red or X-ray flat sources (primarily motivated by the need to eliminate extrinsic dust reddening and gas absorption, respectively) does not pose a significant limitation beyond reducing the final sample size. This strategy directly parallels established low-redshift distance indicators, where specific, highly homogeneous sub-classes of objects are isolated to minimise intrinsic scatter, such as restricting supernova cosmology to normal Type Ia events standardized via light-curve shapes \citep[e.g.,][]{phillips1993, Riess1996, Perlmutter1999}, or limiting stellar distance estimates to specific pulsation modes of Cepheid or RR Lyrae variables \citep[e.g.,][]{Freedman2001}.

\subsection{Slope of the X-ray to UV relation as a function of redshift}
As part of the selection of the AGN sample for cosmology, one needs to verify that the slope of the X-ray to UV relation does not evolve with redshift. 
We have then fitted the $\fx$--$\fo$ relation for the filtered sample of 1995 AGN in narrow redshift bins, with a $\Delta\log (z)$\,=\,0.015. 
To check whether our selection criteria missed out catastrofic outliers, we performed the regression analysis in each redshift interval also applying a sigma-clipping with $\sigma$\,=\,2.6. Only $\sim$\,1\% of the sources (23) are excluded by the clipping, leading to a final sample of 1972 AGN. As this procedure depends upon the binning, we re-fit the data with slightly different intervals and bin width, finding a similar number of outliers ($\pm$\, 2). The main properties of the final SDSS--4XMM AGN sample are listed in Table~\ref{tbl:sample}.

The best-fit slope and intrinsic dispersion values obtained from the regression analysis are plotted in Figure~\ref{fig:slopez}. As evident from this figure, there is no systematic redshift evolution: a linear fit as a function of redshift, of the form $\gamma$\,=\,$mz$\,+\,$q$, returns a flat slope and an intercept consistent with the average $\gamma$ obtained for the AGN sample: $m$\,=\,$-0.02$\,$\pm$\,0.02 and $q$\,=\,0.59\,$\pm$\,0.03 (shown with the grey solid line). 
The dispersion decreases with redshift, a behaviour that is consistent with previous studies in the literature \citep[e.g.,][]{li2021,rankine2024,chira2026}.

\begin{figure}[h]
\centering
\includegraphics[width=\linewidth,clip]{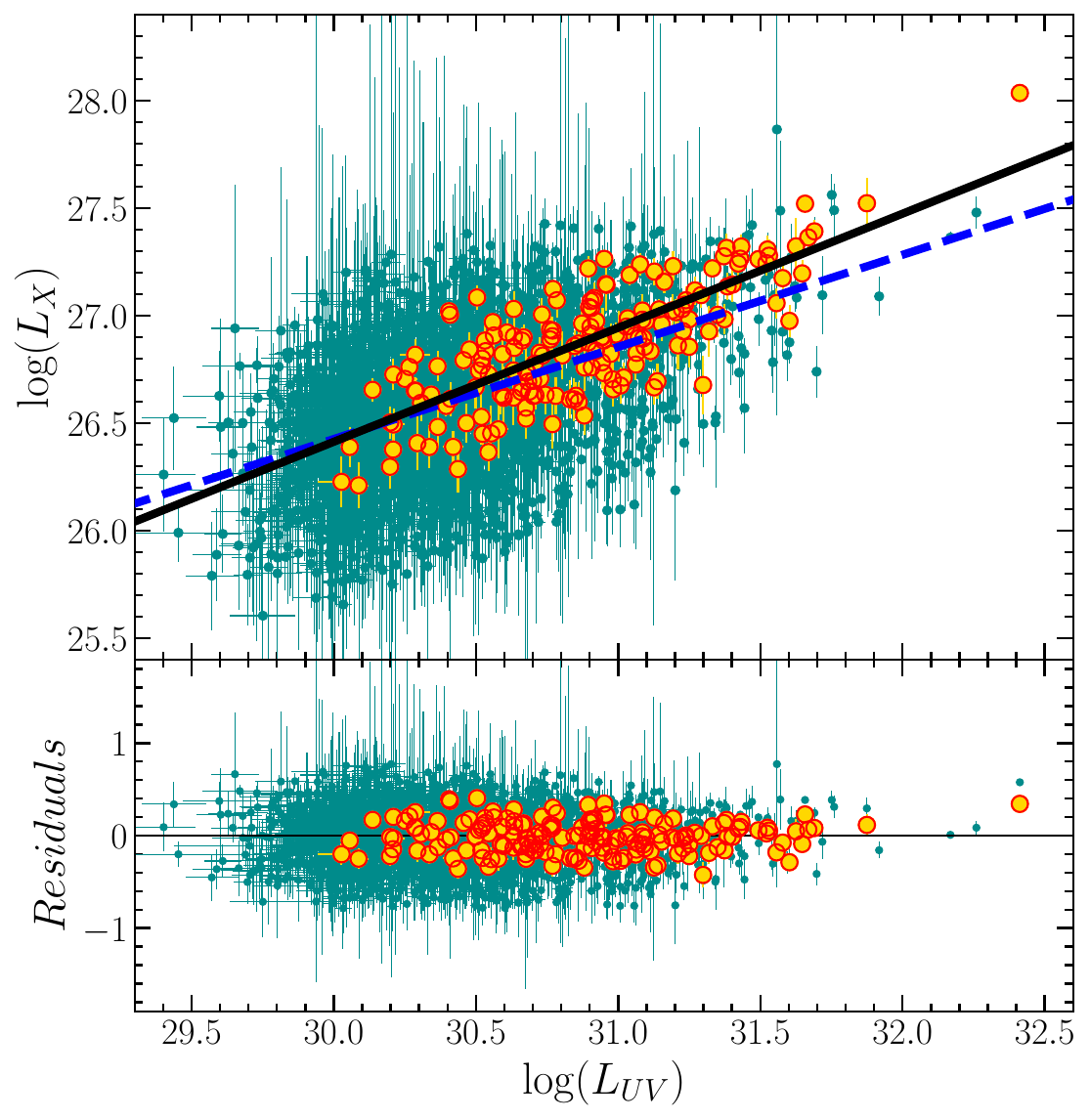}
\caption{X-ray-to-UV relation and its best fit in the redshift interval 2.0--2.5 for the whole sample (teal points, blue dashed line) and for the AGN sample that fulfills the selection criteria described in Section~\ref{Sample Selection} (orange points, black solid line).}
\label{fig:singleintervalz}
\end{figure}
In Figure~\ref{fig:singleintervalz}, we illustrate the explicit impact of our selection criteria on both the slope and intrinsic dispersion of the X-ray-to-UV relation within a representative, narrow redshift interval ($z$\,=\,2--2.4). Prior to applying any filtering, the parent AGN sample within this redshift range exhibits a large dispersion of $\delta$\,$\simeq$\,0.4 dex (teal points), which drastically reduces to $\delta$\,$\simeq$\,0.15 dex once we isolate the clean sub-population that fulfills the selection criteria detailed in Section~\ref{Sample Selection} (orange points). Crucially, the best-fit slope of the relation also shifts significantly, moving from $\gamma$\,=\,0.67 for the unfiltered sample to $\gamma$\,=\,0.60 for the final selection. This local behavior is consistently observed across the entire redshift range probed by our study, demonstrating that the careful mitigation of the Eddington bias is the primary driver behind recovering the true, intrinsic slope of the X-ray-to-UV relation across cosmic time.

\begin{table*}[h!]
\caption{\label{tbl:sample} SDSS--4XMM AGN properties.}
\centering
\begin{tabular}{cccccccc}
\hline\hline
SDSS & $z$ & RA & DEC & $\log\lo$ & $\log\lx$ & $\gammax$ & $N_{\rm obs}$\\
(1) & (2) & (3) & (4) & (5) & (6) & (7) & (8)\\
\hline
  011123.40$+$330225.9 & 0.718 & 17.8477   & 33.0406    & 29.34 $\pm$ 0.03  & 25.96 $\pm$ 0.03 &  2.17 $\pm$ 0.21  & 1\\
  020421.06$-$040818.9 & 0.723 & 31.0877   & $-$4.13867 & 30.26 $\pm$ 0.01  & 26.35 $\pm$ 0.02 &  2.03 $\pm$ 0.11  & 1\\
  021808.24$-$045845.2 & 0.716 & 34.534233 & $-$4.97918 & 30.54 $\pm$ 0.01  & 26.80 $\pm$ 0.04 &  2.20 $\pm$ 0.28  & 3\\
  022617.85$-$043109.1 & 0.708 & 36.5743   & $-$4.51919 & 29.91 $\pm$ 0.01  & 25.93 $\pm$ 0.03 &  2.27 $\pm$ 0.25  & 1\\
  022716.12$-$044539.1 & 0.722 & 36.8172   & $-$4.76086 & 29.56 $\pm$ 0.02  & 26.25 $\pm$ 0.02 &  1.96 $\pm$ 0.10  & 1\\
  022841.08$+$003049.4 & 0.720 & 37.1711   & 0.513728   & 30.43 $\pm$ 0.01  & 26.40 $\pm$ 0.01 &  2.07 $\pm$ 0.10  & 1\\
\hline
\end{tabular}
\tablefoot{Columns: (1) SDSS name; (2) SDSS spectroscopic redshift; (3) and (4) SDSS coordinates; (5) logarithm of $\lo$ (with uncertianties) in units of erg s$^{-1}$ Hz$^{-1}$; (6) logarithm of $\lx$ (with uncertianties) in units of erg s$^{-1}$ Hz$^{-1}$; (7) photon index (from broadband 4XMM fluxes); (8) number of merged XMM-{\it Newton} observations. The entire table is available online.}
\end{table*}

\subsection{X-ray undetected AGN}
Our analysis starts from the cross-match between the SDSS quasar catalogue and the XMM-{\it Newton} catalogue of {\em detected} sources. As such, it does not include possible non-detections in the available fields of view. The presence of a large number of non-detections would significantly bias the results, and requires to include the relative upper limits in the statistical analysis. Considering that, for a typical quasar SED, SDSS is significantly deeper than the available X-ray observations, the issue is possibly critical depending on the amount of missing X-ray detections. 
In general, we expect that our conservative filter on the depth of the X-ray observations solves this problem: by requiring that an observation is deeper by at least 4$\sigma$ than the expected X-ray flux, we should already remove nearly all the non-detections. 
We further checked whether this is the case by applying our filters to the list of cross-matches between the SDSS catalogue and the XMM-{\it Newton} pointings used to produce the 4XMM--DR14 catalogue. 
Adopting a search radius of 12$^{\prime}$ yields 48,790 individual entries, while the total number of matches between the SDSS-DR16 and the 4XMM--DR14 detections is 41,779. Therefore, the initial fraction of non-detections is 14\%. We then applied the following procedure to the two samples:
\begin{itemize}
\item We removed BAL and radio-bright quasars, and we applied the same optical/UV colour filters.
\item We did {\em not} apply any X-ray slope filter, as this is impossible for the non-detections. Therefore, in order to obtain comparable samples, we neglected this filter for the detections as well.
\item We applied the Eddington-bias filter described above. While the reference X-ray flux derived from the UV luminosity is obviously the same in both samples, the estimate of the depth of the X-ray observation is not completely analogous. We formally adoped the same procedure as described above, based on the exposure time and the off-axis angle. However, in the case of the match with the observations pointings list, such a `blind' estimate does not take into account the regions of the detectors where the effective area is much lower than expected, mainly due to (a) the gaps between the chips and some damaged pixel columns in the EPIC detectors, and (b) the failure of two chips of the MOS1 and one in the MOS2 detectors.  According to the {\em XMM-Newton} documentation, the first issue produces a loss of observational area of about 5-8\%, while the second decreases such an area by up to 30\%, depending on the observation date. Considering that the PN instrument is on average deeper than the MOS, we can expect that between 5\% and-10\% of the area encircled in the 12 arcmin radius is not available for scientific observations.
\end{itemize}

The final filtered sample of matches with the {\em XMM-Newton} pointing list consists of 7,385 objects. Considering the loss of detector area discussed above, we conservatively expect that at least 370-400 of these matches are not scientifically useful observations. Therefore, the "true" number of matches will be of the order of, or slightly lower than, 7,000.  

By repeating the same selection on the 4XMM-DR14 catalogue of detections we obtain 7,031 matches. This confirms our initial prediction of no, or very few, non-detections in our final sample.

\section{Impact of Selection Criteria on the $\lx$--$\lo$ Relation}
\label{sec:filter_effects}

\begin{figure*}[t]
\centering
\includegraphics[width=\linewidth,clip]{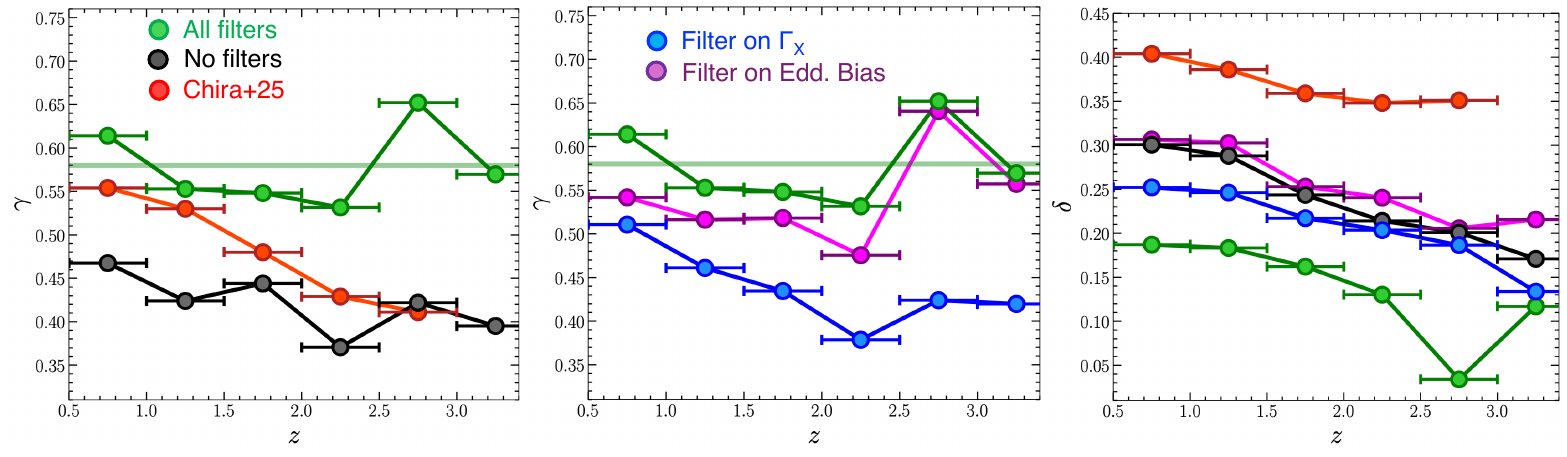}
\caption{Evolution of the $\lx$--$\lo$ relation parameters as a function of redshift in bins of $\Delta z$\,=\,0.5. 
\textit{Left and central panels:} best-fit slope ($\gamma$) for the unfiltered parent sample (black points) and the final selected sample (green points), compared against the results from \citet{chira2026} (red points). The solid horizontal green line denotes the average slope of $\gamma$\,=\,0.58\,$\pm$\,0.01 obtained for our final clean sample (see Figure~\ref{fig:slopez}). 
\textit{Right panel:} intrinsic dispersion ($\delta$) as a function of redshift for the various filtering stages: the unfiltered sample (black); the sample with UV colour and X-ray photon index selection but no Eddington bias correction (blue); the sample with UV colour and Eddington bias correction but no photon index selection (magenta); and the final fully filtered sample (green). The dispersion from \citet{chira2026} is shown in red for comparison.}
\label{fig:slope-chira}
\end{figure*}

In Figure \ref{fig:slope-chira}, we illustrate how successive filtering stages alter both the slope ($\gamma$) and the intrinsic dispersion ($\delta$) of the $\lx$--$\lo$ relation across different cosmic epochs. For comparison, we overlay the recent results from \citet{chira2026}, who detected a statistically significant redshift dependence in the slope ($\gamma$ decreasing at higher redshifts) utilising a large, combined sample of approximately 137,000 AGN with eROSITA and/or XMM-\textit{Newton} coverage.

The physical and statistical drivers of the trend found in our sample prior the various selections, including the Eddington bias one, are clear. In our unfiltered (or partially corrected) data set, the high-redshift domain is heavily shaped by the empirical X-ray detection thresholds. Due to the substantial intrinsic scatter of the whole population (i.e., $>$\,0.35 dex, see Figure~\ref{fig:singleintervalz}), the X-ray flux limit preferentially samples sources undergoing positive statistical fluctuations or those contaminated by additional non-thermal emission mechanisms (e.g., radio jets). This progressive overestimation of X-ray luminosities at higher redshifts artificially flattens the observed $\lx$--$\lo$ slope. Conversely, sequentially introducing our UV colour and X-ray photon index criteria and our conservative Eddington bias threshold systematically isolates a clean, standard accretion-dominated population, simultaneously reducing the intrinsic dispersion (Figure \ref{fig:slope-chira}, right panel) and restoring the true, redshift-independent nature of the disc--corona coupling. Our analysis demonstrates that the observed slope evolution is, at least in our case, a spurious artifact due to the inhomogeneity of the parent sample, in both terms of spectral properties and observational depth. Once the full suite of filters is applied (green points), the slope stabilizes and becomes entirely consistent with a constant value ($\gamma$\,$\simeq$\,0.58) up to $z$\,$\sim$\,3.5.

The divergence between our results and those of \citet{chira2026} can be attributed to two main factors. First, it may stem from an incomplete mitigation of the Eddington bias within the \citet{chira2026} framework. Second, it could reflect a genuine physical distinction, indicating an intrinsic evolutionary trend unique to their broader sample demographics. Disentangling these two scenarios is, however, beyond the immediate scope of the present work, whose primary objective is strictly to isolate a well-characterized population of quasars capable of providing reliable luminosity distance estimates. Ultimately, our filtering framework successfully achieves this goal in either case: if the discrepancy is purely statistical, our approach effectively eradicates a severe observational bias; if the trend is physical, our criteria systematically isolate a homogeneous sub-population governed by a stable, redshift-independent $\lx$--$\lo$ relation.

We note that several different methodological and demographic factors could contribute to the observed divergence between our findings and the evolving trend reported by \citet{chira2026}. First, their sample is predominantly composed of observations from the eROSITA all-sky survey, which, given its relatively shallow flux limit, results in non-detections constituting roughly 95\% of the dataset. Although survival analysis techniques are standard and highly valuable tools for incorporating such upper limits in astronomical surveys \citep[e.g.,][]{Feigelson1985, Isobe1986}, it is well established that the statistical reliability and constraints of these estimators degrade significantly under extreme levels of data censoring \citep[typically exceeding $\sim$80\%; see][]{Feigelson2012}. In a shallow survey layout at high redshifts, the few active detections that anchor the survival analysis models are inherently restricted to the most severe positive statistical fluctuations. 

Consequently, the inclusion of upper limits under these heavily censored conditions may be insufficient to fully decouple the intrinsic relation parameters from the underlying Eddington bias. Furthermore, even assuming a robust statistical treatment of the censoring, an unfiltered sample remains inherently susceptible to cosmic demographic evolution. Because the selection framework in \citet{chira2026} does not implement the strict multi-wavelength filtering utilised in this work, their high-redshift bins are more vulnerable to population variations across cosmic time, such as an evolving fraction of radio-bright or BAL quasars. The preferential detection of intrinsically X-ray-enhanced radio-bright AGN at the lower-UV boundaries of high-redshift bins would naturally introduce a non-physical flattening of the local slope, blending a real demographic shift with severe selection effects.

In addition to the analysis presented in this work (see Fig.~\ref{fig:slope-chira}), a compelling empirical demonstration of the systematic effects introduced by an uncorrected Eddington bias can be found in \citet[Fig.~13]{merloni2024}. In their analysis of the first eROSITA All-Sky Survey Data Release (\text{eRASS1}), the authors compared the X-ray fluxes measured for the same sources detected by both XMM-\textit{Newton} and eROSITA. Because the \text{eRASS1} observations are, on average, significantly shallower than the archival XMM-\textit{Newton} exposures, the measurements at the faint flux regime systematically deviate from a 1:1 relation, with eROSITA reporting systematically brighter fluxes. This discrepancy is the manifestation of the Eddington bias, driven by preferential statistical fluctuations above the shallow survey threshold. When applying an appropriate statistical correction, analogous to the one adopted in this work, the expected 1:1 flux relation is naturally recovered \citep[see Fig.~A.1 of][]{sacchi2025}.

\section{Conclusions}
\label{sec:conclusions}

We have presented a new, highly homogeneous sample of quasars with high-quality X-ray and UV observations, specifically optimised to derive reliable luminosity distances via the non-linear $\lx$--$\lo$ relation. This sample was constructed by cross-matching the Sloan Digital Sky Survey DR16 quasar catalog with the XMM-\textit{Newton} serendipitous source catalogue (4XMM-DR14). To isolate a clean AGN population with uniform properties across both energy regimes, we implemented a rigorous suite of selection criteria:
\begin{enumerate}
    \item constraints on UV and X-ray colours to exclude dust-extincted and gas-absorbed sources;
    \item removal of broad absorption line and radio-bright quasars;
    \item restriction to sources at redshifts $z$\,$<$\,0.7 to mitigate host-galaxy light contamination that could otherwise bias the UV continuum luminosity measurements; and
    \item elimination of sources with shallow X-ray exposures prone to threshold (i.e., Eddington bias) effects.
\end{enumerate}
Our final sample consists of 1,972 clean quasars spanning a wide redshift range ($z$\,=\,0.7--5). We demonstrate that the underlying $\lx$--$\lo$ relation remains remarkably stable across cosmic time, characterised by a constant mean slope of $\gamma$\,=\,0.58\,$\pm$\,0.04 and an intrinsic dispersion of $\delta$\,$\simeq$\,0.15 dex.

The primary objective of this study was to demonstrate that properly correcting for the Eddington bias and selecting a homogeneous sub-class of quasars is absolutely paramount to recovering the intrinsic physics of the accretion disc--corona coupling. At high redshifts, flux limits preferentially catch sources that are statistically brighter than average in the X-rays. Neglecting or under-correcting for this selection effect introduces an artificial, redshift-dependent distortion into the relation parameters. Similarly, a sample containing objects with highly diverse spectral energy distributions can exhibit comparable redshift trends, potentially driven by the distinct evolution of various quasar sub-classes. 

By adopting our selection process, which is completely independent from the observed X-ray to UV ratios, we remove these possible effects and we successfully isolate a homogeneous population governed by a robust, redshift-independent $\lx$$-$$\lo$ relation, consolidating the role of quasars as reliable, independent cosmological probes up to the early Universe.

\begin{acknowledgements} 
MS acknowledges support through the European Space Agency (ESA) Research Fellowship Programme. MR has been supported by the Polish National Agency for Academic Exchange (Bekker grant BPN/BEK/2024/1/00298).

\end{acknowledgements}

\bibliographystyle{aa}
\bibliography{bib}

\appendix
\section{Validation of the X-ray depth filter via simulations}
\begin{figure}[h!]
\centering
\includegraphics[width=0.98\linewidth,clip]{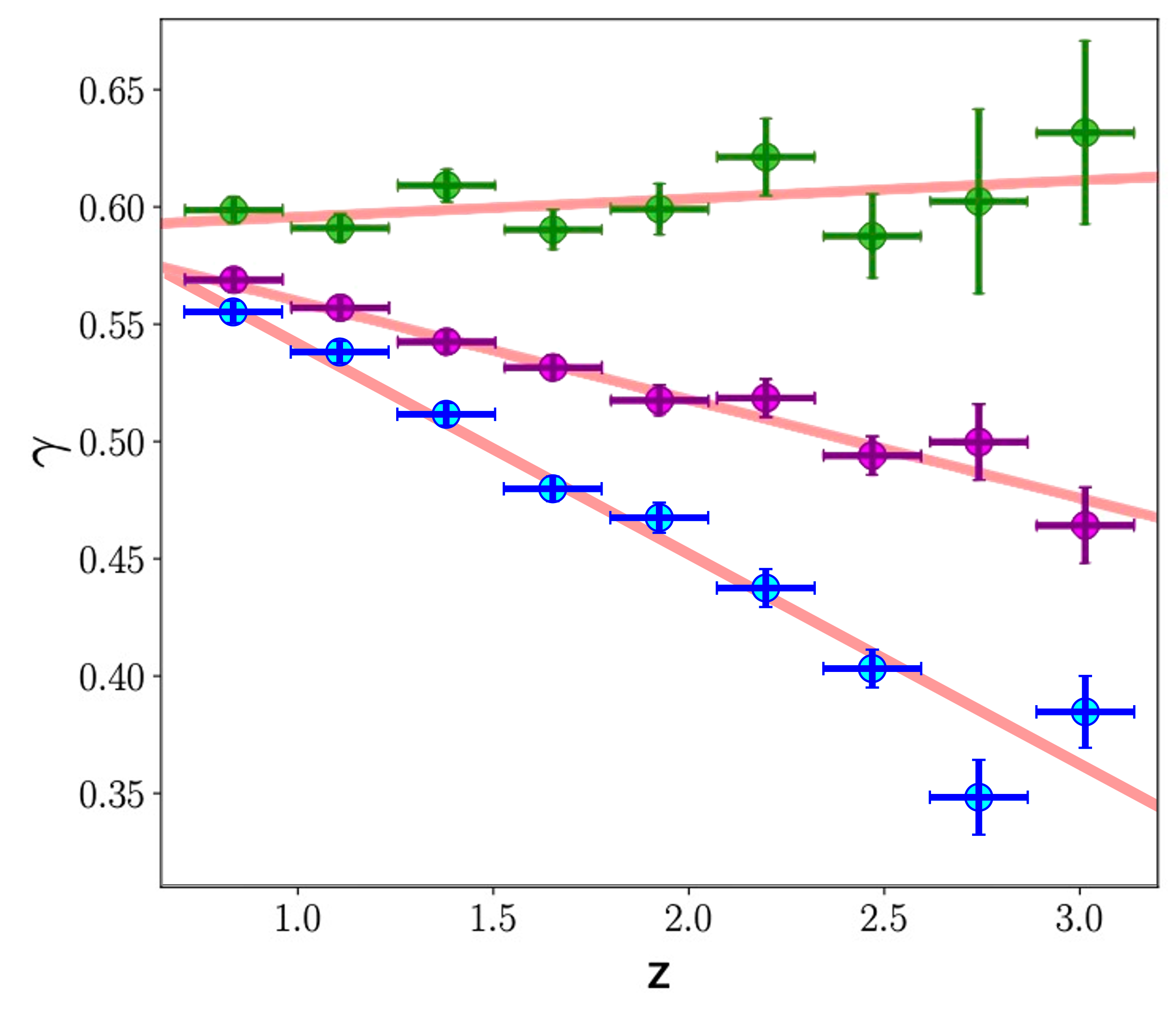}
\caption{Evolution of the $\lx$--$\lo$ relation slope as a function of redshift for the mock samples. The blue and magenta symbols represent the raw simulated data, showing a clear spurious trend with $z$ due to the Eddington bias (selection effects near the flux limit). The blue data points have no correction for the Eddington bias while the magenta points have $K$\,=\,0.35. The green symbols represent the sample after applying the X-ray depth filter ($K$\,=\,0.65). The filtered data correctly recover the input slope (green points) across the probed redshift range, demonstrating the effectiveness of the filter in removing selection-induced biases.}
\label{fig:testlcdm1}
\end{figure}

\begin{figure}[h!]
\centering
\includegraphics[width=0.98\linewidth,clip]{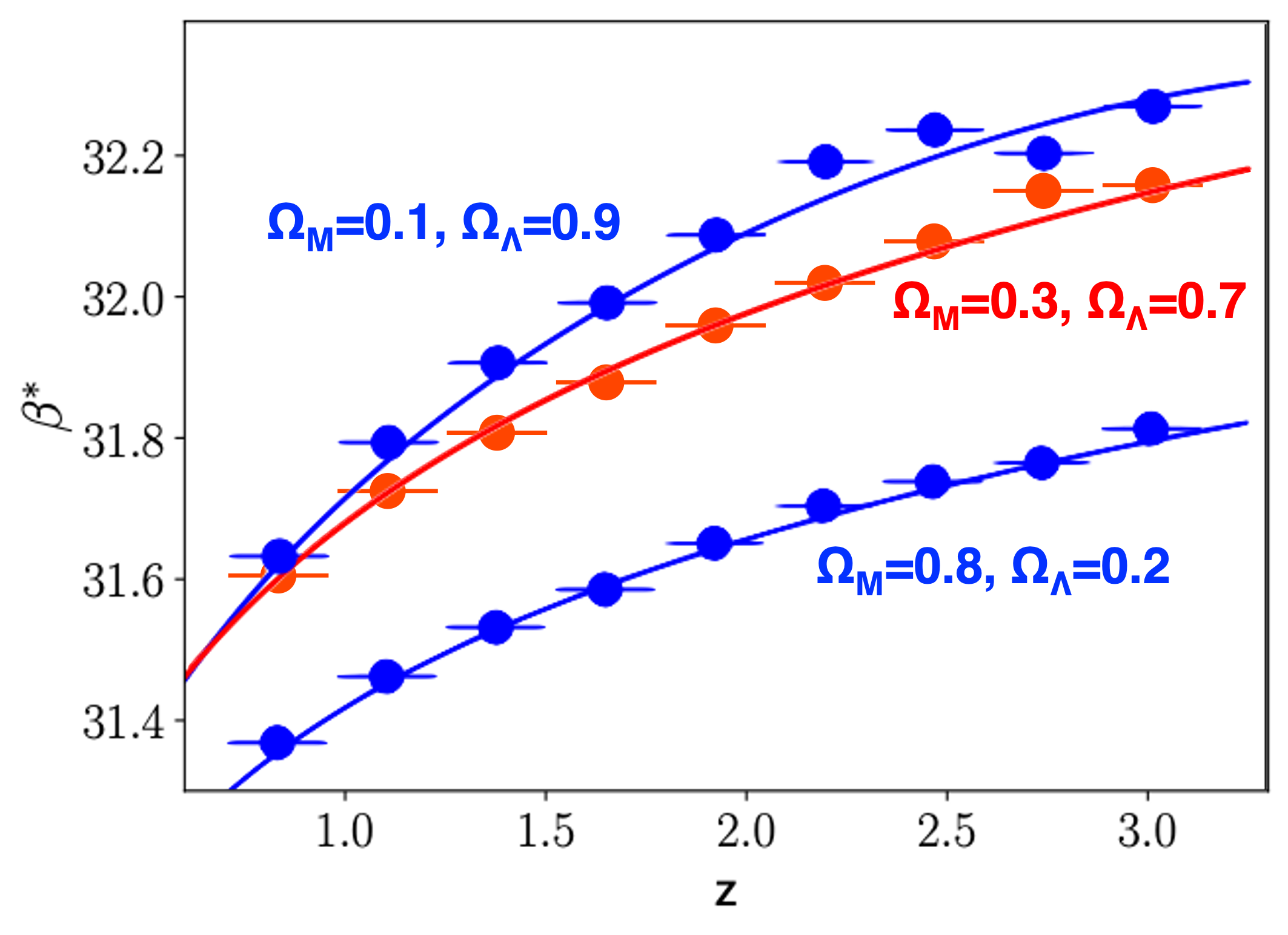}
\caption{Hubble diagrams constructed from the three mock quasar samples simulated with $\om$\,=\,0.1, 0.3, or 0.8. In all cases, the X-ray depth filter was applied assuming a fixed $\om$\,=\,0.3 cosmology. The parameters $\beta^\ast$ represents the intercept of the $\fx$--$\fo$ relation in narrow redshift intervals \citep[see][for details]{lusso2025}. The solid lines represent the best-fit cosmological models, which are in excellent agreement with the true input values. This result confirms that our filtering procedure does not force the data toward the assumed fiducial model but remains conservative enough to recover the underlying `true' cosmology.}
\label{fig:testlcdm2}
\end{figure}

To validate our filtering procedure, we adopted a simulation scheme similar to that discussed in \citet{lusso2025}:
\begin{itemize}
\item We extracted a random sample of 20,000 SDSS quasars at $z$\,$>$\,0.7; this redshift cut was applied to minimize host-galaxy contamination. X-ray fluxes were simulated based on observed UV fluxes by adopting the global X-ray-to-UV luminosity relation. This procedure requires an underlying cosmological model to convert observed UV fluxes into luminosities and subsequently transform the simulated X-ray luminosities back into fluxes. To test the robustness of our method against different cosmological scenarios, we performed three sets of simulations assuming a flat \lcdm\ model with $\om$ values of 0.1, 0.3, or 0.8. These extreme values allow us to verify the ability of our methodology to recover the `true' input model regardless of the initial cosmological assumptions.
\item We introduced a Gaussian dispersion of $\sigma$\,=\,0.15 dex to the simulated X-ray fluxes to account for intrinsic scatter.
\item For each mock sample, we simulated serendipitous XMM-{\em Newton} observations. For every source, a random flux limit was drawn from the distribution of the real XMM-{\em Newton} parent sample. We then assigned flux errors based on the significance of the mock detection or flagged the source as a non-detection if it fells below the assigned limit. The resulting samples yielded detection rates of 80\% for $\om$\,=\,0.8, 90\% for $\om$\,=\,0.3, and 95\% for $\om$\,=\,0.1.
\item We applied the filters described in Section~\ref{Sample Selection}. Crucially, the X-ray depth filter was applied assuming a fixed fiducial model (flat \lcdm\ with $\om$\,=\,0.3) in all three cases, specifically to test if this choice biases the recovery of other cosmologies.
\item Finally, we analysed the X-ray-to-UV relation following the methodology shown in Figure~\ref{fig:slopez} and constructed the quasar Hubble diagrams, which were fitted according to the procedure described in \citet{lusso2025}.
\end{itemize}
The primary results of this analysis, illustrated in Figures~\ref{fig:testlcdm1} and \ref{fig:testlcdm2}, can be summarised as follows:
\begin{enumerate}
\item In the absence of the depth filter, we find a spurious redshift dependence in the slope of the X-ray-to-UV relation (blue points in Fig.~\ref{fig:testlcdm1}). This artifact arises from the Eddington bias, where a higher fraction of X-ray-bright fluctuations are preferentially included at higher redshifts. Once the filter is applied, this dependence vanishes and we successfully recover the redshift-independent slope used in the simulation input.
\item When using the filtered data to derive luminosity distances and construct the Hubble diagrams, the results for all three mock samples are perfectly consistent with their respective input cosmological models. This confirms that while the filtering threshold assumes $\om$\,=\,0.3, the cut is sufficiently conservative to avoid biassing the recovery of any physically motivated cosmological scenario.
\end{enumerate}
As noted in the main text, we repeated this analysis using various selection thresholds. While we successfully recovered the correct cosmology for any $K$\,$>$\,0.5, we maintain a more stringent value of $K$\,=\,0.65 to ensure maximum robustness in our final cosmological constraints.

\end{document}